\documentclass[a4paper,11pt]{article}
\pdfoutput=1 
\usepackage{jcappub}
\usepackage[T1]{fontenc}
\usepackage{appendix}
\usepackage{enumitem}
\usepackage{booktabs}
\usepackage{multirow}
\usepackage{siunitx}
\usepackage{orcidlink}

\title{\boldmath A deep learning approach to cosmological dark energy models}

\author[a]{Celia Escamilla-Rivera\orcidlink{0000-0002-8929-250X},}
\author[b]{Maryi A. Carvajal Quintero}
\author[c]{and Salvatore Capozziello}

\affiliation[a]{Instituto de Ciencias Nucleares, Universidad Nacional Aut\'onoma de M\'exico, Circuito Exterior C.U., A.P. 70-543, M\'exico D.F. 04510, M\'exico.}
\affiliation[b]{Universidad de Antioquia, Calle 70 No. 52-21. Apartado A\'ereo 1226. Antioquia, Colombia.}
\affiliation[c]{Dipartimento di Fisica, University of Napoli ``Federico II'' and Istituto Nazionale di Fisica Nucleare (INFN), Sez. di Napoli, via Cinthia 9, I-80126 Napoli (Italy). Gran Sasso Science Institute (GSSI), via F. Crispi 7, I-67100, L'Aquila (Italy). Laboratory for Theoretical Cosmology, Tomsk State University of Control Systems and Radioelectronics (TUSUR), 634050 Tomsk (Russia).}
\emailAdd{celia.escamilla@nucleares.unam.mx}
\emailAdd{maryi.carvajal@udea.edu.co}
\emailAdd{capozziello@na.infn.it}

\abstract{We propose a novel deep learning tool in order to study the evolution of dark energy models. 
The aim is to combine two architectures: the Recurrent Neural Networks (RNN) and the Bayesian Neural Networks (BNN), we named this full network as RNN+BNN. The first one is capable of learning complex sequential information to classify objects like supernovae and use the light-curves directly to learn information from the sequence of observations. Since RNN is not capable to calculate the uncertainties, BNN emerges as  a solution for problems in deep learning like, for example, the  overfitting. For the trainings we use measurements of the distance modulus $\mu(z)$, such as those provided by Pantheon Supernovae Type Ia. In view of our results, the reported  approach  turns out to be a first promising step on  how we can train a new neural network that can compute their own confidence regions for specific cosmological data. It is worth stressing that the new technique allows  to reduce the computational load of expensive codes for dark energy models and probe the necessity of modified dark energy models at large redshifts for a supernovae trained sampler.}
 
\begin{document}
\maketitle
\flushbottom


\section{Introduction}
\label{sec:intro}

What is dark energy? This question has been addressed in a fundamental context since the discovery of the cosmic acceleration \cite{Riess:1998cb,Perlmutter:1998np} and what can be driving it. Its nature is one of the most fundamental issues facing cosmologists today and it is the subject of several current and future experiments that will survey the sky, e.g DESI \cite{url_desi}, DES \cite{url_des}, WFIRST \cite{url_wfirst} and LSST \cite{url_lsst}. Adjoint to these surveys, current observations as SNeIa, BAO, CMBR anisotropies, LSS formation and 
WL set a strong confirmation of the present-day accelerated expansion of the universe,  consistent with the current standard cosmological model $\Lambda$CDM, where the $\Lambda$ is usually related to an extra constant component  of cosmological fluid (with equation of state --EoS-- $w=-1$).  Such a cosmological constant, if evolving, is dubbed  dark energy.  However, we cannot measure  directly dark energy because we  can only observe its  effects  on the Hubble flow measuring   observable components like  matter and radiation. Since the standard model interplays gravity and dark energy, we can encode the distribution of matter in the universe in $w=p(\rho)$, where $p$ is the pressure and $\rho$ the matter-energy density. As a simple choice, we can assume $w=p/\rho$. It  can be characterised by a distribution derived from  statistical measures in two (or three)-point correlation functions \cite{Takada:2002qq,Marin:2013bbb} or other frequentist statistics \cite{Tsujikawa:2010sc,Huterer:2000mj,Wang:2008zh,EscamillaRivera:2011qb,Bull:2015stt}. 

Currently, a great debate  is around the validity of $\Lambda$CDM  model, where tensions are arising between Planck \cite{Aghanim:2018eyx} and other cosmological measurements as: Cefeids (SH0ES), strong lensing time delays (H0LiCOW), tip of the red giant branch  (TRGB), megamasers, Oxygen-rich Miras and surface brightness fluctuations \cite{Verde:2019ivm}. These tensions which justify the study of possible alternatives to the concordance model. One of the most interesting approach seeks for  dynamical properties of dark energy, which should be able to mimic $\Lambda$ at the present time as required by the cosmological observations. The approach starts from quintessence scenarios \cite{Ratra:1987rm,ArmendarizPicon:2000dh}, dark energy parameterisations \cite{Sendra:2011pt, Zhao:2015wqa,Escamilla-Rivera:2016qwv,Rezaei:2017yyj,Escamilla-Rivera:2019aol}, modify gravity \cite{Jaime:2013zwa,Lazkoz:2018aqk}, extended theories of gravity \cite{Capozziello:2019cav},  $w(z)$ reconstructions \cite{Vazquez:2012ce}, non-parametric reconstructions of $w(z)$ \cite{seikel, Montiel:2014fpa}, till Bayesian reconstruction of a time-dependent EoS \cite{Zhao:2017cud} or dynamical $w_{x}$ from $f(R)$ models \cite{CapozzielloQC,Jaime:2018ftn} 
which offers a large overview on how we are trying to explain the effects of dark energy. However, a consensus of a unique model is still missing and all the proposals imply a dependency which can be significantly different by imposing a different theoretical landscape. Clearly, current and future data from the surveys will certainly clarify the issue by improving the determination of $H_0$ and $w(z)$, meanwhile we are dealing with wide distributions of matter proposals.

In order to present a relevant step in the study of dark energy, one can use a technique  capable to combine all features in the distribution of matter that can give an insight into the nature of dark energy. A powerful technique is the deep learning (DL), which is a field of machine learning (ML) that uses several layers of nonlinear processing neurons to obtain and transform at each successive layer an output from the previous layer \cite{Aurelien}.  
The algorithms for this subset are created and operate similar to those in ML, but there are numerous layers of these algorithms, each providing a different interpretation to the data it feeds on. 
Many problems in modern cosmological data analysis are tightly related to fundamental problems in ML, such as classifying stars and galaxies and cluster finding of dense galaxy populations \cite{Charnock:2016ifh,Mathuriya:2018luj}. Other typical problems include data reduction, probability density estimation, how to deal with missing data and how to combine data from different surveys \cite{Kessler:2010wk,Moss:2018tug,Moss:2019fyi}. Unfortunately, up to now, there is not target distribution of $w(z)$ for dark energy  where one  has an efficient parameterisation with an error parameter space. 
An increasing part of modern cosmology aims at the development of new statistical data analysis tools and the study of their behaviour and systematics is often not aware of recent developments in ML and computational statistics. 
Of course, with the emergence of big data in recent years, the DL architecture will optimise algorithms that can handle the complexity of new cosmological models. 

Here we present a novel technique to study dark energy models using a DL trained network in order to identify theoretical cosmological models through the data processed within layers of the network. In comparison to ML, DL technique does not require any labeled/structure data, as it relies on the different outputs processed by each layer which is amalgamated to form an unified way to identify theoretical models via images (matrices of observational data). Another novel result of our proposal is the ability to compute the confidence regions for each kind of training, within so far in ML astrophysical methods this has not been employed before. For this purpose,  we combine the architecture of Recurrent Neural Networks (RNN) with the architecture of Bayesian Neural Networks (BNN) in view of a join network RNN+BNN representation learning, which can yield  parameter spaces for the cosmological models. The advantage of this idea is that the proposed RNN+BNN network is trained and tested on observed supernovae data, and otherwise has no physical parameterisation related to the physical processes of cosmic acceleration. This is different in comparison if we train simulated data, e.g CMB and gravitational lensing, where a physical parameterisation needs to be used.

This paper is organised as follow: in Sec.\ref{sec:background1} we describe four cosmological models: two standard ones and two unified dark fluid equations of state, in order to point out the issues to achieve the cosmological analysis. In Sec.\ref{sec:pantheon} we denote the astrophysical sampler used. In Sec.\ref{sec:DL} we explain the basics of the deep learning technique employed to design a standard RNN architecture and how to add the Bayesian architecture in order to solve problems like overfitting. In Sec.\ref{sec:RNN-BNN} we present the methodology to train the observational supernovae data using four activation functions on our proposed neural network RNN+BNN.
In Sec.\ref{sec:bayesian_DL} we perform a bayesian comparison on the robustness of our RNN+BNN network approach against the proposed cosmologies, and use a $\Lambda$CDM simple scenario to demonstrate how to carry out parameter estimation from our networks results. Finally in Sec.\ref{sec:conclusions} we discuss our former results.


\section{Models for dark energy equations of state}
\label{sec:background1}

In this paper we are going to consider four dark energy models: from the standard scenarios as $\Lambda$CDM and CPL to unified dark energy fluid scenarios as Generalised Chaplygin gas (GCG) and Modified Chaplygin gas (MCG) models. In the following lines we enlisted these dark energy models with their respectively mathematical derivation.

$\Lambda$CDM is a Friedmann-Robertson-Walker (FRW)  cosmology with a partition $\rho$ into (at least) three
components: matter $\rho_m$, radiation $\rho_r$ and a poorly understood dark energy $\rho_{\Lambda}$,
where the latter goes one step further by also invoking the constraint $w=-1/3$. Towards this direction, many
attempts have already been done, starting from the above simple relation to several complex ones, always an explicit relation between
the pressure $p$ and density $\rho$.
\begin{itemize}
\item $\Lambda$CDM model. We shall take the standard model: $H(z)^2/H_0^2 =\Omega_m (1+z)^3 + (1-\Omega_m)$,
where  $w_{\Lambda}=-1$, which provide
a good fit for a large number of observational data compilations
without addressing some important theoretical problems, such as the cosmic coincidence and the fine tuning of the $\Lambda$ value \cite{Peebles:2002gy}.

\item CPL model.
So-called by Chevallier-Polarsky-Linder \cite{Chevallier:2000qy,Linder:2007wa}, can be
represented by two parameters that exhibit the present value of the EoS $w_0$ and its overall time evolution $w_a$:
\begin{eqnarray}
 w (z)_{\text{CPL}}= w_0 +\left(\frac{z}{1+z}\right) w_a.  
\end{eqnarray}
\end{itemize}

%
As an extrapolation of these standard scenarios, 
in the same homogeneous and isotropic universe framework,
we consider that the gravitational sector
to which the matter sector is minimally coupled, 
is described by the standard General Relativity. We also assume that the total energy of the universe comes in the form of photons ($\gamma$), baryons (b), neutrinos ($\nu$) and a unified dark fluid (UDF, $X_U$) \cite{Paul:2013sha}. This component can behave as dark energy, dark matter or a different type of fluid while the universe expands. Therefore, the full energy budget is denoted by $p_{i}/\rho_{i}$, where $i=\gamma, b, \nu, X_U$.  
Additionally, each fluid $i$ obeys a continuity equation of the  form
$\dot{\rho}_i + 3 \frac{\dot{a}}{a} (p_i + \rho_i) =  0.$
%
Standard solutions are: $\rho_b \propto a^{-3}$ and $\rho_{\gamma,\nu} \propto a^{-4}$. For a UDF, a constant adiabatic sound speed $c_{s}$ is assumed and can be modelled as $p=c_{s}^2 (\rho-\tilde{\rho})$, where $c_{s}$ and $\tilde{\rho}$ are positive constants. One part of this expression behaves as the usual barotropic cosmic fluid and the other as a $\Lambda$, which unifies the dark energy and dark matter components (effect so-called as dark degeneracy). By integrating for a UDF we obtain:
\begin{eqnarray}
\rho = \rho_{\Lambda} + \rho_{X_U} a^{-3(1+c_s^2)}, \\
p= -\rho_{\Lambda} + c_s^2 \rho_{X_U} a^{-3(1+c_s^2)},
\end{eqnarray}
where $\rho_{\Lambda}=c_s^2 \tilde{\rho}/(1+c_s^2)$ and $\rho_{X_U}=\rho_0 -\rho_{\Lambda}$, with $\rho_0$ as the dark energy density at the present time. The
dynamical EoS is given by:
\begin{equation}
w= -1 +\frac{1+c_s^2}{\left(\frac{\rho_\Lambda}{\rho_{X_U}}\right)(1+z)^{-3(1+c_s^2)}+1}.
\end{equation}
At this point, a specific form which specifies $p_{X_U}$ as a function of $\rho_{X_U}$: $p_{X_U} = f(\rho_{X_U})$ needs to be considered:

\begin{itemize}
\item Generalised Chaplygin gas (GCG) model.
For a given $X_U$ is characterised by:
$p_{\rm gcg} = -\frac{A}{(\rho_{\rm gcg})^{\alpha}}\,,$
%
where $A$ and $0\leq \alpha \leq 1$ are two free parameters. The case $\alpha= 1$ correspond to the original Chaplygin gas model. Solving the continuity equation using the evolution of the GCG energy density, one gets:
\begin{equation}
\rho_{\rm gcg}(a)=\rho_{{\rm gcg},0} \left [ b_{s}+(1-b_{s})a^{-3(1+\alpha)} \right ]\,
^{\frac{1}{1+\alpha}},
\label{eq:rhogcg}
\end{equation}
where $\rho_{{\rm gcg},0}$ denotes the energy density of the GCG fluid at present time and $b_{s}=A\rho_{{\rm gcg},0}^{-(1+\alpha)}$. This density contains all the $i$ components described above. Then we can compute the GCG dynamical equation of state $w_{\rm gcg}$ as a function of the redshift $z$:
\begin{eqnarray}
w_{\rm gcg}(z)=-\frac{b_{s}}{b_{s}+(1-b_{s})\left(\frac{1}{1+z}\right)^{-3(1+\alpha)}}.
\label{eq:eosgcg}
\end{eqnarray}

This establish the regions of dominations for an effective dark matter component and an effective dark energy one, with an intermediate region for $\alpha=1$. 


\item Modified Chaplygin gas (MCG) model.
Its relation between pressure $p_{\rm mcg}$ and energy density $\rho_{\rm mcg}$  is given by:
%
$p_{\rm mcg} =  B \rho_{\rm mcg} - \frac{A}{(\rho_{\rm mcg})^{\alpha}}\,,$
%
where again $A$, $B$, and $\alpha $ are three real constants with $0\leq \alpha \leq 1$. If $A=0$, the MCG behaves as a perfect fluid with $w=B$, whereas, if $B=0$, we can recover the GCG model. Again, the standard Chaplygin gas model can be obtained by setting $\alpha=0$. Solving  the equation we find that the evolution of the MCG energy density with all its $i$ components:
\begin{equation}
\rho_{\rm mcg}(a)=\rho _{{\rm mcg},0} \left [ B_{s}+(1-B_{s})a^{-3(1+B)(1+\alpha)} \right ]^{\frac{1}{1+\alpha}},
\label{eq:rhomcg}
\end{equation}
where $\rho_{{\rm mcg},0}$ denotes the energy density of the MCG fluid at present time, and $B_{s}=A\rho_{{\rm gcg},0}^{-(1+\alpha)}/(1+B)$. The evolution of the MCG EoS is given by:
\begin{equation}
w_{\rm mcg}(z)=B-\frac{B_{s}(1+B)}{B_{s}+(1-B_{s})\left(\frac{1}{1+z}\right)^{-3(1+B)(1+\alpha)}}.
\label{eq:eosmcg}
\end{equation}

As an extension of the GCG model, the MCG model behaves accordingly with each cosmological regions described.
\end{itemize}


\section{Pantheon Type Ia supernovae compilation}
\label{sec:pantheon}

Since we are interested in the behaviour of the cosmic late acceleration using the proposed dark energy models, in this paper we employ the recent Type Ia supernovae (SNeIa) sample called Pantheon  \cite{Scolnic:2017caz}. 

This sampler consists of 1048 SNeIa in 40 bins compressed. It is the
largest spectroscopically confirmed SNeIa sample to date. This characteristic makes it attractive to develop with our DL test.
Since we are performing tests of EoS's that, at some point, recover $\Lambda$CDM, the binned catalog is not a problem in the sense of favoring this model.

SNeIa can give determinations of the distance modulus $\mu$, whose theoretical prediction is related to the luminosity distance $d_L$ according to:
\begin{equation}\label{eq:lum}
\mu(z)= 5\log{\left[\frac{d_L (z)}{1 \text{Mpc}}\right]} +25,
\end{equation}
where the luminosity distance is given in Mpc. In the standard statistical analysis, one adds to the distance modulus the nuisance parameter $M$, an unknown offset sum of the supernovae absolute magnitude (and other possible systematics), which is degenerate with $H_0$. 

As we are assuming spatial flatness, the luminosity distance is related to the comoving distance $D$ via
$d_{L} (z) =\frac{c}{H_0} (1+z)D(z),$
where $c$ is the speed of light, so that, 
we can obtain
\begin{equation}
D(z) =\frac{H_0}{c}(1+z)^{-1}10^{\frac{\mu(z)}{5}-5}.
\end{equation}
Therefore, the normalised Hubble function $H(z)/H_0$ can be obtained by taking the inverse of the derivative of $D(z)$ with respect to the redshift 
\begin{equation}
D(z)=\int^{z}_{0} H_0 d\tilde{z}/H(\tilde{z}), \label{eq:dist}
\end{equation}
where $H_0$ is the Hubble constant consider as a prior value to normalise $D(z)$.



\section{Basics on Deep Learning} 
\label{sec:DL}
The deep learning technique allows algorithms to learn more complex patterns by adding more layers to the Neural Network (NN), using Convolutional (CNN) or Recurrent Neural Networks (RNN) \cite{Ntampaka:2019udw}.
DL has demonstrated its versatility in the application of new methods where rich data (as the supernovae Pantheon sampler presented above) will play an important role. Additionally, it allows to automatically built a (usually highly nonlinear) model that maps from a given input (e.g a redshift $z$) to an output (e.g. a modulus distance $\mu(z)$) with different algorithms to use several prescriptions to build up a specific model.

Can this information be used by the DL algorithm to gain a better understanding of a dark energy cosmological model? 
Currently, the interest over these kind of algorithms are bringing new opportunities for data-driven cosmological discovery, but they will also present new challenges for adopting DL methodologies and understanding the results when the data are too complex for traditional model development and fitting with statistics. Some proposals in that regards have been done in order to explore the DL methods to measure cosmological parameters for future large-scale photometric surveys \cite{Charnock:2016ifh} and from density fields \cite{Schmelzle:2017vwd}.
Even more, the advantage to use DL is that the overfitting can be avoided, it increases the robustness of classifier and carries a coverage of trained data. In our case, we select the extensive knowledge of supernovae to obtain a DL training data sampler. 

In this paper we used, as first piece of the new network proposed, a supervised learning training called Recurrent Neural Network (RNN), which is a type of algorithm that adopts the real target to teach the NN and penalizes when it is far from the real data. Then this algorithm output is compared with the using of a loss function, i.e. we take into account the Mean Squared Error (MSE) function, where the objective of the algorithm is to minimise the loss function. Afterwards, we use the Adam optimizer \cite{ruder} to find the minimum and obtain the best cosmological model. In order to compute the confidence errors for our cosmological models, we unify this first piece of architecture with our second piece, a Bayesian Neural Network (BNN). Along the rest of this paper we refer to the union of these architectures as our RNN+BNN network.

In the next two subsections, we describe each architecture separately. Afterwards, we will explain how to join both in order to design our RNN+BNN network.


\subsection{Training Recurrent Neural Networks (RNN)} 
\label{ssec:RNN}

There are many applications of DL for large photometric surveys, such as: (1) the measurement of galaxy shapes from images; (2) automated strong lens identification from multi-band images; (3) automated classification of supernovae; (4) galaxy cluster identification. We will focus on a non-linear regression-like method for supernovae using deep RNN. It was show in \cite{Charnock:2016ifh,moller} that deep RNN are capable to learn complex sequential information  to classify supernovae. In our approach, instead of performing feature extraction before classification, we use the redshifts of light-curves directly as inputs to a RNN, which is able to learn information from the sequence of observations.
Due to the intrinsic relation of our data with time, we employ RNN that has show a successful behaviour with sequential series, being used in problems like translation or transcription. These kind of networks are similar to Feed Forward Neural Networks (FFNN) \cite{Aurelien,Goodfellow}, but have an essential difference in comparison to RNN since a recurrent network feed itself during the training, e.g. one network with a neuron will have a connection from the input and also from the output of its previous time step --from which feed itself-- a characteristic which make them adequate to train a supernovae data sampler. We will explain mathematically this aspect in the next system of equations. 

Even more, these connections allow the algorithm to have a contest above all the sequence, alike FFNN where we must  show all the sequence at the same time, so, these networks could not identify who is first and who is not. Then, the FFNN only see a gate as a free parameter which it does not feed itself with the latter value, i.e there is not a direct connection with the backward data.

\begin{figure}[h]
    \centering
    \includegraphics[width=0.75\textwidth,origin=c,angle=0]{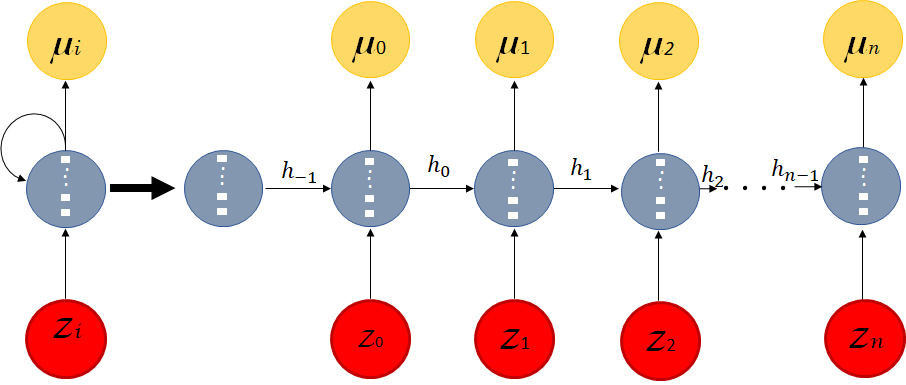}
    \caption{RNN architecture. The input $z_i$ are the redshifts and the output $y_i$ are the corresponding modulus distance $\mu_i$. This network example has one layer, and the small white squares inside each circle node represent the number of neurons $h_{i}$. In our proposal the number of neurons is equal to 100. 
    Each neuron receives inputs from other neurons through the paths. After several steps there is some processing on the input and the calculated 
    result is passed to other neurons connected through another path. Each node is associated to an activation function (Tanh, ReLU, ELU and SELU) which transforms the input to an output value. The output value of these functions acts as an input to the next connected neurons. Final output helps in deciding the class of input data. The number of steps is given by $n$.
    }
    \label{fig:RNN_arc}
\end{figure}

To design a RNN architecture, we start with a cell were the output of the previous time step is used to compute the new one (see Figure \ref{fig:RNN_arc}). The same weights are used to compute all the sequence, and each cell is feeding with the output before using the following expression: 
\begin{eqnarray}
    h^{<t>}&=&g(W_{h}\cdot h^{<t-1>} + W_{x}\cdot x^{<t>} +b_{a}),\\
    y^{<t>}&=&g(W_{y}\cdot h^{<t>} +b_{y}),
\end{eqnarray}
where $ h^{<t>}$ is called the hidden state and its  $ h^{<t-1>}$ value represent the hidden state before it, $g$ is the activation function and $y^{<t>}$ is the output. Here,  $t$ is the time referring  to the first sequence ($t=1$, for the first redshift) with the data and $b$ are the bias. From now on, the superscript $<  >$ indicates a vector and the dot is a matrix product.

However, basic RNN fails in large sequences due to the loss of information from the initial inputs. In order to improve our training performance we use a modification of RNN cells, the so-called Long Short Term Memory (LSTM) cells \cite{Aurelien}. The goal of these kind of cells are their capability of forgetting and adding information step by step. However,  a lot of modifications have been done to these cells by incrementing the number of matrix functions for each layer which correspond to \cite{Zaremba}:
\begin{eqnarray}
    i^{<t>}&=&\sigma(W_{xi}^{T}\cdot x^{<t>}+W_{hi}^{T}\cdot h^{<t-1>}+b_{i}),\label{eq:input} \\
    f^{<t>}&=&\sigma(W_{xf}^{T}\cdot x^{<t>}+W_{hf}^{T}\cdot h^{<t-1>}+b_{f}), \label{eq:forget}\\
    o^{<t>}&=&\sigma(W_{xo}^{T}\cdot x^{<t>}+W_{ho}^{T}\cdot h^{<t-1>}+b_{o}), \label{eq:output}\\
    g^{<t>}&=&A_{f}(W_{xg}^{T}\cdot x^{<t>}+W_{hg}^{T}\cdot h^{<t-1>}+b_{g}), \label{eq:newstate}\\
    c^{<t>}&=&f^{<t>}\otimes c^{<t-1>}+i^{<t>}\otimes g^{<t>}, \label{eq:finalstate}\\
    y^{<t>}&=&h^{<t>}=o^{<t>}\otimes A_{f}(c^{<t>}) \label{eq:outputcell},
\end{eqnarray}
where $W$ are the weights of each layer, $\sigma$ is the sigmoid function that takes values between 0 and 1. Here $\otimes$ is the direct product. In the rest of the paper, we denote the superscript $T$ as the transpose of the quantity where it is indicated. Notice how the feed process is explicitly given in the second r.h.s term of the Eqs. (\ref{eq:input})-(\ref{eq:newstate}).

We compare several trainings for $\mu(z)$ using four kind of activation functions $A_f$ defined as\footnote{We consider here the values for $\alpha =1.673$ and $\lambda= 1.051$ for a monotic behaviour.}\cite{monti}:
\begin{eqnarray}   
A_{f_{\text{Tanh}}} &=& \tanh(x), \quad \text{in} \quad (-1,1), \label{eq:tanh} \\
A_{f_{\text{ReLU}}} &=& \left\{ \begin{array}{ll}
         0 & \mbox{for $x \leq 0$},\\
        x & \mbox{for $x > 0$}.\end{array} \right.  \quad \text{in} \quad [0,\infty), \label{eq:relu} \\
A_{f_{\text{ELU}}} &=&    \left\{ \begin{array}{ll}
         \alpha (e^x -1) & \mbox{for $x \leq 0$},\\
        x & \mbox{for $x > 0$}, \end{array} \right.  \text{in} \quad (-\alpha,\infty),  \label{eq:felu} \\      
 A_{f_{\text{SELU}}} &=&     \left\{ \begin{array}{ll}
         \alpha \lambda (e^x -1) & \mbox{for $x \leq 0$},\\
        x & \mbox{for $x > 0$}, \end{array} \right.  \text{in} \quad (-\alpha\lambda,\infty).  \label{eq:selu} \quad \quad
\end{eqnarray}

In this full system of equations, we consider three layers called \textit{gate controllers}:
    (a) \textit{Input gate:} This gate is giving by (\ref{eq:input}) and helps  the cell to know which values must remember in a long term.
    (b)
    \textit{Forget gate:} Giving by (\ref{eq:forget}), this gate allows  the cells to forget certain values. 
    (c) \textit{Output gate:} It controls which values will be the output and which not, as it is seen in (\ref{eq:output}).
Eq.(\ref{eq:newstate}) gives a possible new state to remember in a long term. However, according to the LSTM, the final state (\ref{eq:finalstate}) and (\ref{eq:outputcell}) give the output of the cell, e.g if we obtain a null vector for the forget gate, taking into account (\ref{eq:finalstate}), the network will not remember anything about the previous state. Whereas, if we obtain values of one, it will remember everything.

Nevertheless, NN have a large number of parameters, which can cause a high possibility of overfitting. In addition of these technical parameters, we have at hand the free parameters from each cosmological models. Therefore, we 
improve this by using a \textit{regularization} technique \cite{Goodfellow}. But, in RNN few regularization models works well and a good suggestion is the \textit{Variational Dropout} (VD) \cite{DropoutRNN}. 
\textit{Dropout} is a powerful and inexpensive method of regularization, which randomly turns off some neurons to avoid overfitting and the probability of turning off a neuron is given by the hyperparameters. These can be adjusted depending on size, epochs, layers, neurons and batchsize adopted. 
In the VD,  we repeat the same dropout mask for  inputs, hidden states and outputs. This method allows a good behaviour for dropout in RNN and 
is very useful in BNN to compute the corresponding confidence contours for our trainings.

\subsection{Bayesian Neural Networks (BNN)} They emerge as  a solution for problems in DL like overfitting and the incapability of networks to calculate their uncertainty.
Uncertainties are crucial to evaluate the validity of a reliable result. So far, for astrophysical scenarios, the error calculations are misleading due the deterministic performing of the networks and at this point they are unable to provide a good estimation of the error.

Therefore, we propose a combination of RNN and bayesian approaches by BNN that can solve this problem by making the networks probabilistic. We positioned them in order to deal with the uncertainty issue and take all assumptions into account, i.e. we can obtain a probability distribution over the weights, given the distributional training data, to be used as next inputs. At the end of the process, we are capable to obtain an entire distribution over the network output, which increases our predicting accuracy and confidence regions. For this purpose, we need to
know what kind of distribution follows the weights. Having a prior distribution on their weights and their biases, and giving a dataset \textbf{X, Y}, we search for the posterior distribution over the space of parameters: $p(\omega| X, Y)$. Hence,  we can predict an output for a new input point $x$ by integrating the following:
\begin{equation}
    p(\mathbf{y}^{*}|\mathbf{x}^{*},\mathbf{X},\mathbf{Y})=\int p(\mathbf{y}^{*}|\mathbf{x}^{*},\mathbf{\omega})p(\mathbf{\omega}|\mathbf{X},\mathbf{Y})\mathbf{d\omega},
\end{equation}
This \textit{dropout} could be used in model uncertainty estimation and it is an approximation of the Gaussian process. In this way, we can calculate model uncertainty applying dropout in the training $n$-times. We consider a BNN modelled with a simple dropout which offers a same mask at each time step leading to improved results and without overfit quickly \cite{DropoutRNN,DLmatrix}.


\section{RNN+BNN architecture for deep learning dark energy EoS} 
\label{sec:RNN-BNN}

In this section we describe the methodology to train the observational data given by the Pantheon supernovae sample in Sec.\ref{sec:pantheon} using our proposed RNN+BNN network. We analise four different architectures of activation functions. The implementation of these analyses were done using TENSORFLOW \cite{TF}. All of them were built out of layers shown in Figure \ref{fig:RNN_arc}. 
The first layer of the network is a RNN with one activation function. The last layer is a connected layer with no activation function to map the output value of the last residual neuron to the desired output size. The value of the size needs to be fixed at hand in order to reduce the dispersion of the data at low redshifts while the NN adjusts during the training. 
The control of this value can be achieved by computing the $\chi^2$ for each activation function.  
The RNN+BNN network was trained on two cosmological parameters: an input $z$ and the output $\mu$. The advantage of this process if that we can add a BNN to the architecture to always predict a valid covariance matrix. Finally, we used a negative log-likelihood loss function
\begin{equation}
L= \frac{1}{2}\left(\ln{\zeta + (\theta_{p}-\theta_t)^{T}\zeta^{-1}(\theta_{p}-\theta_t) }\right),
\end{equation}
where $\theta_p$ if the vector of predicted parameters, $\zeta^{-1}$ is the predicted inverse covariance matrix given by BNN and $\theta_t$ are the input (real) parameters. 
It is important to remark that at this first stage of the DL technique we can only use data that share the same cosmological observations, e.g. luminosity (modulus) distance. While the results of several data samplers are already statistically promising, we found that combining the prediction of different datasets have high noise levels which make difficult the convergence of our RNN+BNN network.

\begin{figure*}
    \centering
      \includegraphics[width=0.38\textwidth,origin=c,angle=0]{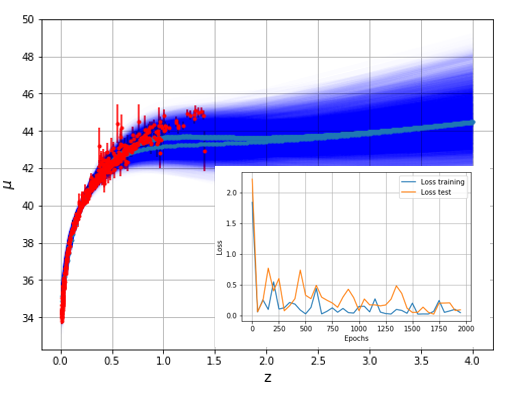} \quad
 \includegraphics[width=0.38\textwidth,origin=c,angle=0]{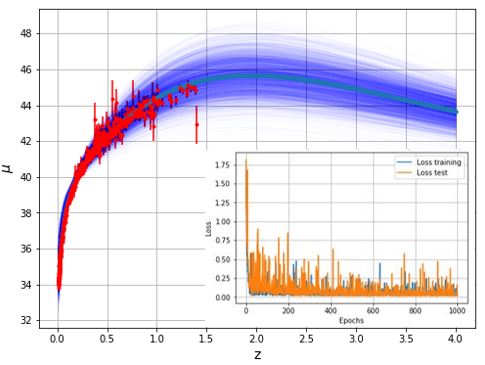}
 \includegraphics[width=0.38\textwidth,origin=c,angle=0]{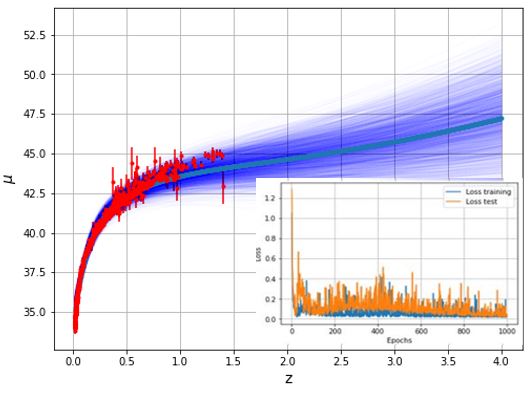} \quad
 \includegraphics[width=0.38\textwidth,origin=c,angle=0]{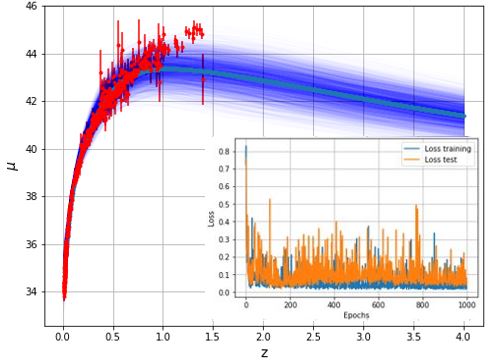}
    \caption{RNN-BNN network training models for Pantheon sampler using activation functions (\ref{eq:tanh})-(\ref{eq:selu}) from top (left) to bottom (right), respectively. \textit{Outside plot:} The uncertainty contours at 1-2$\sigma$ are represented by blue color, the green line represent the mean data trained and the red color dots are the real data. \textit{Inside plot:} \textit{Loss} plots were also created to show the mean absolute error loss over the training epochs for both the train (blue color) and test (orange color) sets.}
    \label{fig:trainings}
\end{figure*}

\begin{figure}
    \centering
    \includegraphics[width=0.48\textwidth,origin=c,angle=0]{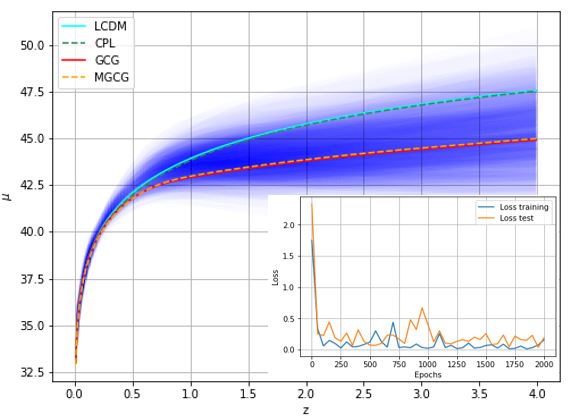} \quad
    \includegraphics[width=0.5\textwidth,origin=c,angle=0]{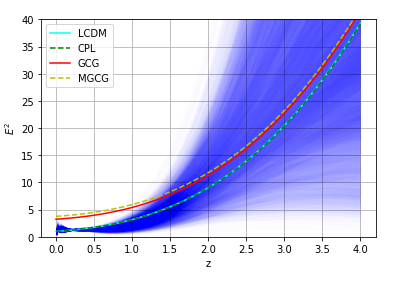}
    \caption{\textit{Top:} Trainings for dark energy EoS: $\Lambda$CDM model (blue line), CPL model (green-dashed line), GCG model (red line) and MGCG (yellow-dashed line) using the activation function (\ref{eq:tanh}). \textit{Bottom:} Trainings for $E^2(z)$ for each model.}
    \label{fig:de_EoS}
\end{figure}

In order to construct and train our network, we carry out the following algorithm:
\begin{enumerate}[label=(\roman*)]
        \item Construction of the neural network. For the RNN, our model uses the LSTM cell described in Sec.\ref{ssec:RNN}. We consider one layer with 100 neurons and use a projector to obtain the same shape of the input. The input is $z$ and the output $\mu$. We construct the sequence from lower to higher redshift $z$. 
          \item Organising the cosmological data. After ordering from lower to higher redshift $z$, we need to re-arrange the data using the number of steps $n$. According to our architecture in Fig. \ref{fig:RNN_arc}, we choose a number of steps $n=4$.
        \item Computing the BNN. 
        Our dropout has the following parameter: the probability to drop the input is $0$. This is because, after testing it several times, we found that our models could not be trained with an input dropout due the lost of information. Alternatively, the state dropout is $0.4$ and the output is $0.5$ \cite{Aurelien}. Finally, the cost function is MSE type, therefore we can use the Adam optimizer.
        \item Training the RNN+BNN neural network. We start the training by consider $1000$ epochs. Afterwards, we read the model and apply $500$ times the same dropout to our initial model. This allows to obtain the uncertainty contours (Figure \ref{fig:trainings}). After the training, we predicted a data sampler of 1000 data points.
               \item Calculation of the bestfits. We use the CLASS 
        \footnote{\url{https://github.com/lesgourg/class_public}} 
        and Monte Python codes
        \footnote{\url{https://github.com/baudren/montepython_public}} 
        to constrain the models using the new Pantheon sampler obtained from the RNN+BNN training (Step v). For each theoretical model proposed we obtained the values reported in Tables \ref{tab:lcdm}-\ref{tab:cpl}-\ref{tab:gcg}-\ref{tab:mcg}.
                 \item Calculation of $\mu(z)$. For each of our cosmological models, we consider now $2000$ epochs to redo (iv). Once we obtain the new predicted sampler, we perform the integration in (\ref{eq:dist}) to compute the functions $\mu(z)$ and $E^2(z)$ by using a specific dark energy EoS in terms of $z$. The results for each model are given in Figure \ref{fig:de_EoS}.
                 

            \end{enumerate}
 
Given the stochastic nature of our training algorithm for Pantheon sampler, some specific results may vary. We run our chain up to 100 times in order to obtain a homogeneous result and to extrapolate the redshift up to $z=4$, given as a result that the model learned the problem, achieving a near zero error, at least to two decimal places. In Figure \ref{fig:trainings}, it is worth noticing that for each activation function the training models seems to have converged. The loss plots show good convergence behaviour, although somewhat bumpy. The model may be good configured given no sign of over or under fitting and the learning rate or batch size may be tuned to even out the smoothness of the convergence in this case.
We change the activation functions to improve our performance and reduce uncertainty contours. As it is expected, uncertainty contours change with the activation functions due the change on the covariance matrix. As a main result, the activation function (\ref{eq:tanh}) gives the better evolution to perform our dark energy trainings. This result will be probed from a bayesian point of view in the next section.  
According to the trainings (Figure \ref{fig:de_EoS}), we notice that unified dark fluids models seems to be preferred at large redshift less than $1$-$\sigma$ in comparison to the standard models.

The models in Figure \ref{fig:trainings} are in agreement of what it is expected of the activation functions: ReLU is not implemented in models with one layer and neither in LSTM cells since their unbounded values could easily explode.  LSTM cells commonly use the sigmoid and Tanh activation functions. Our results support the use of this last activation function in these types of cells. 

Our proposed RNN+BNN network algorithm for SNeIa is composed by 8 hyperparameters: 
\begin{itemize}
\item Size=4, Epochs=1000, Layers=1, Neurons=100, Bathsize=10.
\item A variational dropout composed by: an input $z$, a hidden state $h$ and an output $\mu(z)$.
\end{itemize}
As a first tentative exploration, these DL parameters can be interpreted as the price we pay in order to deal with a model-independent dark energy cosmology. Moreover, a homogeneous sampler as SNeIa is not a complex model \footnote{More complex models, e.g natural language, requires thousands of inhomogeneous data.}, therefore it only needs one layer as much, since we do not require greater depth in the model. If we add more layers the computational cost is unnecessary and can lead us to overfitting. Also, we notice that the number of neurons is high, but it is important to remark that the dropout controls any overfitting that this can carried out. Each of the latter hyperparameters can add complexity to the model, which can lead us to overfitting (or underfitting), but none of them --at least for this homogeneous SNeIa sampler)-- can change drastically the covariance matrices, as the $A_f$ can indeed do it \cite{DropoutRNN}. 


\section{Bayesian versus deep learning significance} 
\label{sec:bayesian_DL}

In this section we present a Bayesian model selection to describe the relationship between the cosmological models described above and their DL trainings, the supernovae data and the prior information about the free parameters. \mbox{Using Bayes} theorem \cite{bayes-th}
we can updated the prior model probability to the posterior model probability. Additionally, when we compare
models, the evidence can be used to evaluate the model's evolution using the sampler available. The evidence can be denoted by
\begin{eqnarray}\label{eq:bayes}
\mathcal{E} =\int{\mathcal{L}(\theta) P(\theta) d\theta},
\end{eqnarray}
where $\theta$ is the vector of free cosmological parameters for each model and $P(\theta)$
is the prior distribution of these parameters. The evidence can be computed with several methods \cite{gregory,Trotta:2005ar}, but for our purpose we applied a  multinested sampling algorithm \cite{skilling} which has proven practicable in cosmology applications \cite{Liddle:2006kn}.

\subsection{Information criteria results}

Since we are working with four cosmological models with different number of cosmological parameters, the simplest procedure to compare them is the likelihood ratio test, which can 
be applied when the simple model is nested within a more complex model \cite{Liddle:2004nh}. The 
quantity $2\ln{\cal L}_{{\rm simple}}/{\cal L}_{{\rm complex}}$, where ${\cal 
L}$ is the maximum likelihood 
of the model under consideration, is approximately chi-squared distributed and Jeffreys's scale
can be used to look up the significance of any 
increase in likelihood against the number of extra parameters introduced. Also, due the number of free cosmological parameters and data sampler, we consider two statistical criterias in order to compare our DL architecture results (Tables \ref{table_af_lcdm}-\ref{table_af_cpl}-\ref{table_af_gcg}-\ref{table_af_mcg}) with the observational ones (Table \ref{table_bestfits1}):

The Akaike information criterion (AIC) is defined as
\begin{equation} \label{eq:aic}
\mathrm{AIC} = -2\ln {\cal L} + 2k\,,
\end{equation}
where ${\cal L}$ is the maximum likelihood and 
$k$ the number of parameters of the model. The best model is
the model which minimizes the AIC.
Usually models with few parameters 
give a poor fit to the data and hence have a low log-likelihood, while those 
with many are penalized by the second term in (\ref{eq:aic}). 

The Bayesian information criterion (BIC) is defined as
\begin{equation}\label{eq:bic}
\mathrm{BIC} = -2\ln{\cal L} + k \ln N \,,
\end{equation}
where $N$ is the number of data points used in the fit.

{\renewcommand{\tabcolsep}{10.mm}
{\renewcommand{\arraystretch}{0.6}
\begin{table*}
\begin{minipage}{\textwidth}
\caption{Results for the dark energy parameterisations with best fits parameter using observational Pantheon sampler. In 3-column it is show the $\chi_{min}^2$ for each model and in 4-column the likelihood for each model. } 
\resizebox*{\textwidth}{!}{
\begin{tabular}{|c|c|c|c|c|}
\hline {\bf Model}        &{\bf Parameterisation }    &  $\chi_{min}^2$ &$-\ln{\cal L}_\mathrm{min}$ \\
\hline\hline $\Lambda$CDM & $ w=-1 $    &$1027$ &$513.430$        \\
\hline CPL & $ w (z)= w_0 +\left(\frac{z}{1+z}\right) w_a.  $ & $1028$ &$514.237$ \\
\hline GCG & $w(z)=-\frac{b_{s}}{b_{s}+(1-b_{s})\left(\frac{1}{1+z}\right)^{-3(1+\alpha)}}$ & $1050$ &$525.085$\\
\hline MCG & $w(z)=B-\frac{B_{s}(1+B)}{B_{s}+(1-B_{s})\left(\frac{1}{1+z}\right)^{-3(1+B)(1+\alpha_0)}}$ & $1028$ &$514.248$ \\
\hline 
\end{tabular}\label{table_bestfits1}}
\end{minipage}
\end{table*}}}

{\renewcommand{\tabcolsep}{12.mm}
{\renewcommand{\arraystretch}{0.6}
\begin{table*}
\begin{minipage}{\textwidth}
\caption{$\chi^2$ analyses for $\Lambda$CDM using the activations functions (\ref{eq:tanh}), (\ref{eq:relu}), (\ref{eq:felu}) and (\ref{eq:selu}). } 
\resizebox*{\textwidth}{!}{
\begin{tabular}{ |c|c|c|c| } 
\hline
Model & Activation function & $\chi_{min}^2$ \\
\hline\hline
\multirow{4}{5em}{\centering $\Lambda$CDM} 
& $A_{f_{\text{Tanh}}} = \tanh(x)$ & 606 \\ 
& $A_{f_{\text{ReLU}}} = \left\{ \begin{array}{ll}
         0 & \mbox{for $x \leq 0$},\\
        x & \mbox{for $x > 0$}.\end{array} \right. $  & 647 \\ 
& $A_{f_{\text{ELU}}} =    \left\{ \begin{array}{ll}
         \alpha (e^x -1) & \mbox{for $x \leq 0$},\\
        x & \mbox{for $x > 0$}, \end{array} \right. $ & 570 \\ 
& $ A_{f_{\text{SELU}}} =     \left\{ \begin{array}{ll}
         \alpha \lambda (e^x -1) & \mbox{for $x \leq 0$},\\
        x & \mbox{for $x > 0$}, \end{array} \right. $ & 408 \\
\hline
\end{tabular}\label{table_af_lcdm}}
\end{minipage}
\end{table*}}}

{\renewcommand{\tabcolsep}{12.mm}
{\renewcommand{\arraystretch}{0.6}
\begin{table*}
\begin{minipage}{\textwidth}
\caption{$\chi^2$ analyses for CPL using the activations functions (\ref{eq:tanh}), (\ref{eq:relu}), (\ref{eq:felu}) and (\ref{eq:selu}). } 
\resizebox*{\textwidth}{!}{
\begin{tabular}{ |c|c|c|c| } 
\hline
Model & Activation function & $\chi_{min}^2$ \\
\hline\hline
\multirow{4}{5em}{\centering CPL} 
& $A_{f_{\text{Tanh}}} = \tanh(x)$ & 932 \\ 
& $A_{f_{\text{ReLU}}} = \left\{ \begin{array}{ll}
         0 & \mbox{for $x \leq 0$},\\
        x & \mbox{for $x > 0$}.\end{array} \right. $  & 971 \\ 
& $A_{f_{\text{ELU}}} =    \left\{ \begin{array}{ll}
         \alpha (e^x -1) & \mbox{for $x \leq 0$},\\
        x & \mbox{for $x > 0$}, \end{array} \right. $ & 741 \\ 
& $ A_{f_{\text{SELU}}} =     \left\{ \begin{array}{ll}
         \alpha \lambda (e^x -1) & \mbox{for $x \leq 0$},\\
        x & \mbox{for $x > 0$}, \end{array} \right. $ &  559 \\
\hline
\end{tabular}\label{table_af_cpl}}
\end{minipage}
\end{table*}}}

{\renewcommand{\tabcolsep}{12.mm}
{\renewcommand{\arraystretch}{0.6}
\begin{table*}
\begin{minipage}{\textwidth}
\caption{$\chi^2$ analyses for GCG using the activations functions (\ref{eq:tanh}), (\ref{eq:relu}), (\ref{eq:felu}) and (\ref{eq:selu}). } 
\resizebox*{\textwidth}{!}{
\begin{tabular}{ |c|c|c|c| } 
\hline
Model & Activation function & $\chi_{min}^2$ \\
\hline\hline
\multirow{4}{5em}{\centering GCG} 
& $A_{f_{\text{Tanh}}} = \tanh(x)$ & 1314 \\ 
& $A_{f_{\text{ReLU}}} = \left\{ \begin{array}{ll}
         0 & \mbox{for $x \leq 0$},\\
        x & \mbox{for $x > 0$}.\end{array} \right. $  & 1367 \\ 
& $A_{f_{\text{ELU}}} =    \left\{ \begin{array}{ll}
         \alpha (e^x -1) & \mbox{for $x \leq 0$},\\
        x & \mbox{for $x > 0$}, \end{array} \right. $ & 1055 \\ 
& $ A_{f_{\text{SELU}}} =     \left\{ \begin{array}{ll}
         \alpha \lambda (e^x -1) & \mbox{for $x \leq 0$},\\
        x & \mbox{for $x > 0$}, \end{array} \right. $ & 753 \\
\hline
\end{tabular}\label{table_af_gcg}}
\end{minipage}
\end{table*}}}

{\renewcommand{\tabcolsep}{12.mm}
{\renewcommand{\arraystretch}{0.6}
\begin{table*}
\begin{minipage}{\textwidth}
\caption{$\chi^2$ analyses for MCG using the activations functions (\ref{eq:tanh}), (\ref{eq:relu}), (\ref{eq:felu}) and (\ref{eq:selu}). } 
\resizebox*{\textwidth}{!}{
\begin{tabular}{ |c|c|c|c| } 
\hline
Model & Activation function & $\chi_{min}^2$ \\
\hline\hline
\multirow{4}{5em}{\centering MCG} 
& $A_{f_{\text{Tanh}}} = \tanh(x)$ & 4243\\ 
& $A_{f_{\text{ReLU}}} = \left\{ \begin{array}{ll}
         0 & \mbox{for $x \leq 0$},\\
        x & \mbox{for $x > 0$}.\end{array} \right. $  & 1651\\ 
& $A_{f_{\text{ELU}}} =    \left\{ \begin{array}{ll}
         \alpha (e^x -1) & \mbox{for $x \leq 0$},\\
        x & \mbox{for $x > 0$}, \end{array} \right. $ & 1180 \\ 
& $ A_{f_{\text{SELU}}} =     \left\{ \begin{array}{ll}
         \alpha \lambda (e^x -1) & \mbox{for $x \leq 0$},\\
        x & \mbox{for $x > 0$}, \end{array} \right. $ & 922 \\
\hline
\end{tabular}\label{table_af_mcg}}
\end{minipage}
\end{table*}}}

{\renewcommand{\tabcolsep}{0.5mm}
{\renewcommand{\arraystretch}{0.6}
\begin{table*}
\begin{minipage}{\textwidth}
\caption{Bayesian Deep Learning Pantheon results for the dark energy parameterisations using the the activations functions (\ref{eq:tanh}), (\ref{eq:relu}), (\ref{eq:felu}) and (\ref{eq:selu}). In each the subcolumn it is show the Bayes factor $\ln B_{ij}$, AIC (\ref{eq:aic}) and BIC (\ref{eq:bic}), respectively. Each value is compared with the pivot $\Lambda$CDM model using the trained RNN+BNN.} 
\resizebox*{\textwidth}{!}{
 \begin{tabular}{lSSSSSSSSSSSS}
    \toprule
    \multirow{3}{*}{Model} &
      \multicolumn{3}{c}{$A_{f_{\text{Tanh}}}$} &
      \multicolumn{3}{c}{$A_{f_{\text{ReLU}}} $} &
       \multicolumn{3}{c}{$A_{f_{\text{ELU}}}$} &
      \multicolumn{3}{c}{ $ A_{f_{\text{SELU}}}$} \\ \\
      & {$\ln B_{ij}$} & { {\bf AIC} } & { {\bf BIC} }  & {$\ln B_{ij}$} & { {\bf AIC} } & { {\bf BIC} } & {$\ln B_{ij}$} & { {\bf AIC} }& { {\bf BIC} } & {$\ln B_{ij}$} & { {\bf AIC} }& { {\bf BIC} } \\
      \midrule
    CPL & 5.825 & 957.29 & 940 & 5.819 & 996.29 & 979 & 5.213 & 766.29 & 749 & 5.098 & 584.29 &  567\\ \\
    \hline \\
    GCG & 6.580 & 1339.29& 1322 & 6.597 & 1392.29 & 1375 & 6.209 & 1080.29 & 1063 & 5.879 & 778.29 & 761 \\ \\
    \hline \\
    MCG & 8.204 & 1682.61 & 1661  & 6.930 & 4274.61 & 4253 & 6.444 & 1211.61 & 1190 & 6.278 & 953.61 & 932\\ \\
    \bottomrule
  \end{tabular}\label{table_bayesiancom}}
\end{minipage}
\end{table*}}}

\subsection{Evidence criteria results}

We compute the logarithm of the Bayes factor between two models $\mathcal{B}_{ij}=\mathcal{E}_{i}/\mathcal{E}_{j}$,
where the reference model ($\mathcal{E}_{i}$) with highest evidence is the $\Lambda$CDM model. To compare between $\Lambda$CDM and the other models we use Jeffreys's scale \cite{jeffreys}: if
$\ln{B_{ij}}<1$ there is not significant preference for the model with the highest evidence; if $1<\ln{B_{ij}}<2.5$ the
preference is substantial; if $2.5<\ln{B_{ij}}<5$ it is strong; if $\ln{B_{ij}}>5$ it is decisive. 
\begin{table*}
\begin{center}
\begin{tabular}{|c|c|c|}
\hline $\ln{B_{i0}}$   		    & Strength of evidence    		&  Color region        \\
\hline >5	    			& Strong evidence for model $i$	 &Yellow    				 \\
\hline [2.5,5]            		& Moderate evidence for model $i$ 	  & Red   					 \\
\hline [1,2.5] 		    	& Weak evidence for model $i$ 	    &Blue 	\\
\hline [-1,1]  			& Inconclusive 	   				&Green \\
\hline 
\end{tabular}
\caption{Jeffreys's scale. The Bayes factor of CPL, GCG and MCG parameterisation were computed in comparison to the model $0$, $\Lambda$CDM. We used the information from Table \ref{table_bayesiancom}.}
\label{tab:jefreys}
\end{center}
\end{table*}

\begin{figure*}
    \centering
    \includegraphics[width=0.67\textwidth,origin=c,angle=0]{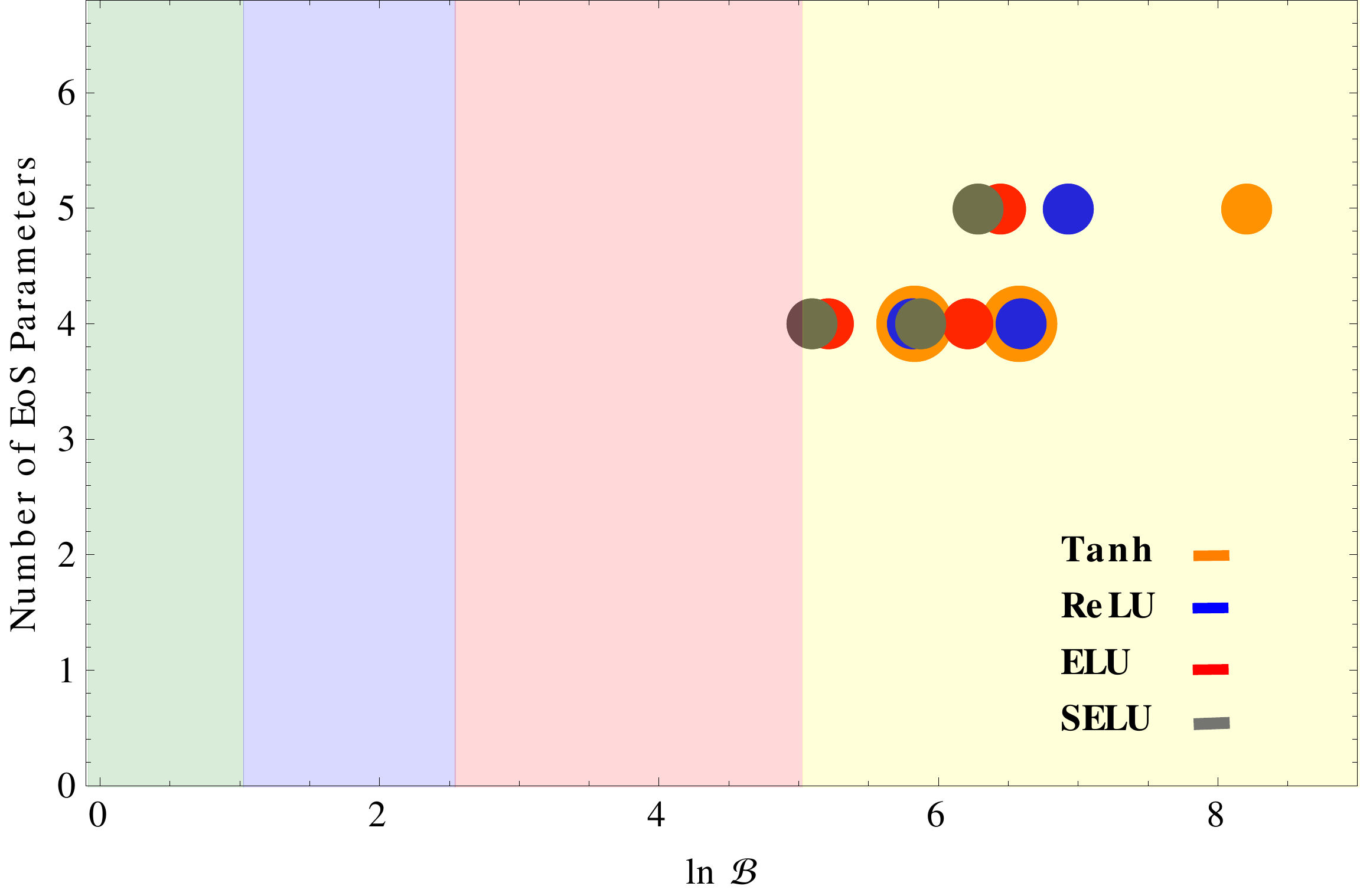}
          \caption{Bayesian portrait according to the number of free EoS parameters using the values given in Table \ref{table_bayesiancom}. The qualitative color regions are detailed in Table \ref{tab:jefreys}. Each color in the plot legend indicates the
activation function used to calculated the value $\ln B_{ij}$ in comparison to $\Lambda$CDM model. The MCG model is represented at level $\nu=5$ and CPL and GCG model at $\nu=4$. }
    \label{fig:bayesian_comp}
\end{figure*}

The results of these analyses are reported in Table \ref{table_bayesiancom}. We compute AIC and BIC statistical criteria for each activation function and cosmological model. Each value is compared with the pivot $\Lambda$CDM  model using the trained RNN+BNN. From here we notice that the activation function Tanh is statistically better in comparison to the others. Also, in order to set a conclusive result in regards to this activation function we design a Bayesian portrait (see Figure \ref{fig:bayesian_comp}) according to the number of free EoS parameters using the values given in Table \ref{table_bayesiancom}. The qualitative color regions are detailed according to the Jeffreys's scale (Table \ref{tab:jefreys}). We denote the number of free cosmological parameters for each model in the y-axes and the evidence in x-axes. From Figure \ref{fig:bayesian_comp} the yellow region indicates that if the models fall there, it means that the trainings used for these models have a statistical preference over $\Lambda$CDM. All our models are in this region. The more advanced towards the yellow region are the models, a strongest evidence against  $\Lambda$CDM appears. Notice that the MCG model using a Tanh activation function is the best of all. ReLU and Tanh functions compete with each other (the blue and red circles overlap for CPL model and GCG model), but for MCG model with a ReLU activation function shows a better evidence in comparison to ELU function. SELU is definitely far behind in this scale.

\subsection{Statistical predictions using RNN+BNN network results}
\begin{figure}
    \centering
    \includegraphics[width=0.5\textwidth,origin=c,angle=0]{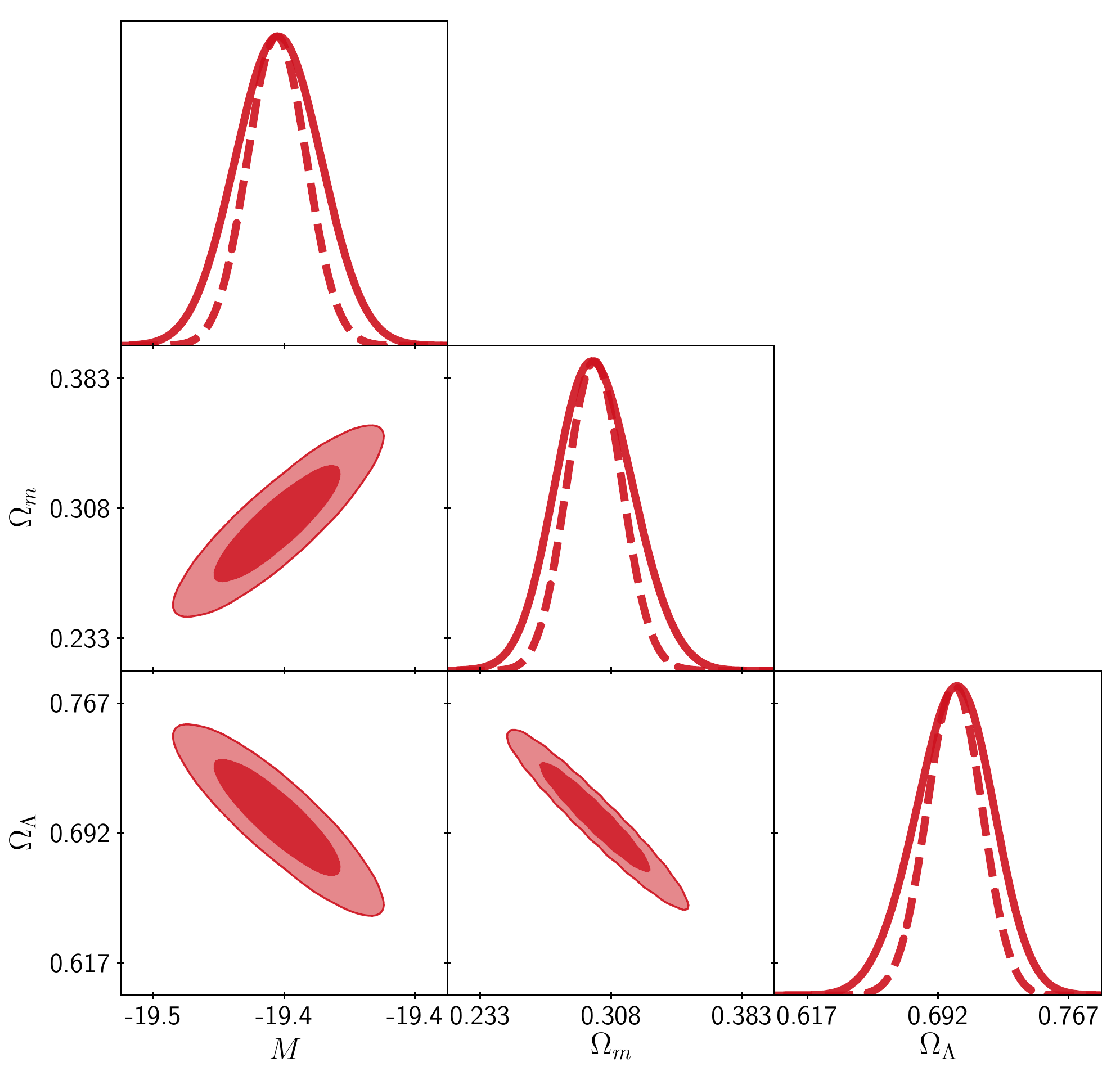}
           \includegraphics[width=0.45\textwidth,origin=c,angle=0]{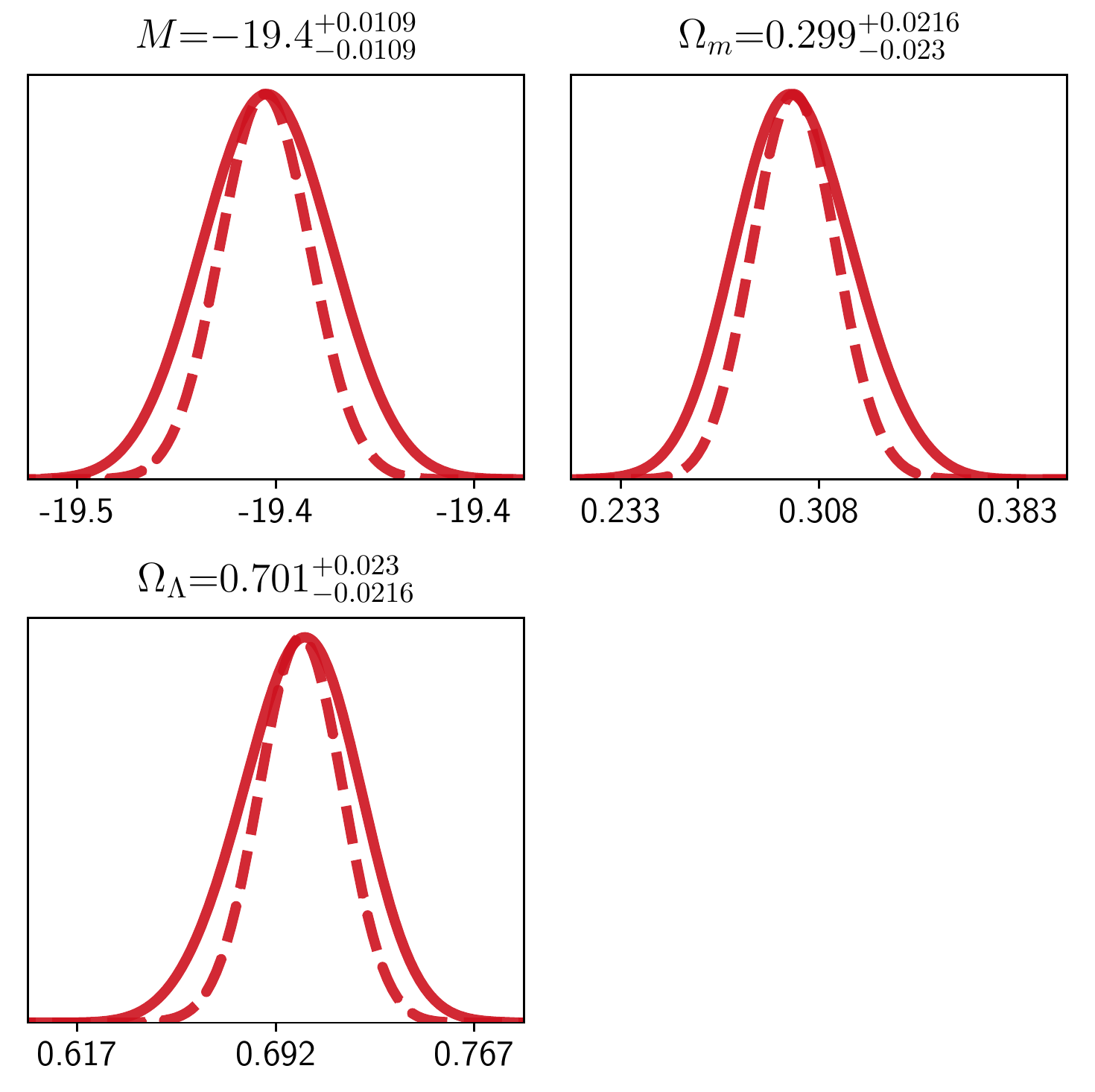}
    \caption{\textit{Left:} Confidence contours, \textit{Right:} Posterior, for $\Lambda$CDM model using Deep Learning Pantheon sampler.}
    \label{fig:bayesian_pantheon_lcdm}
\end{figure}

\begin{figure}
    \centering
    \includegraphics[width=0.57\textwidth,origin=c,angle=0]{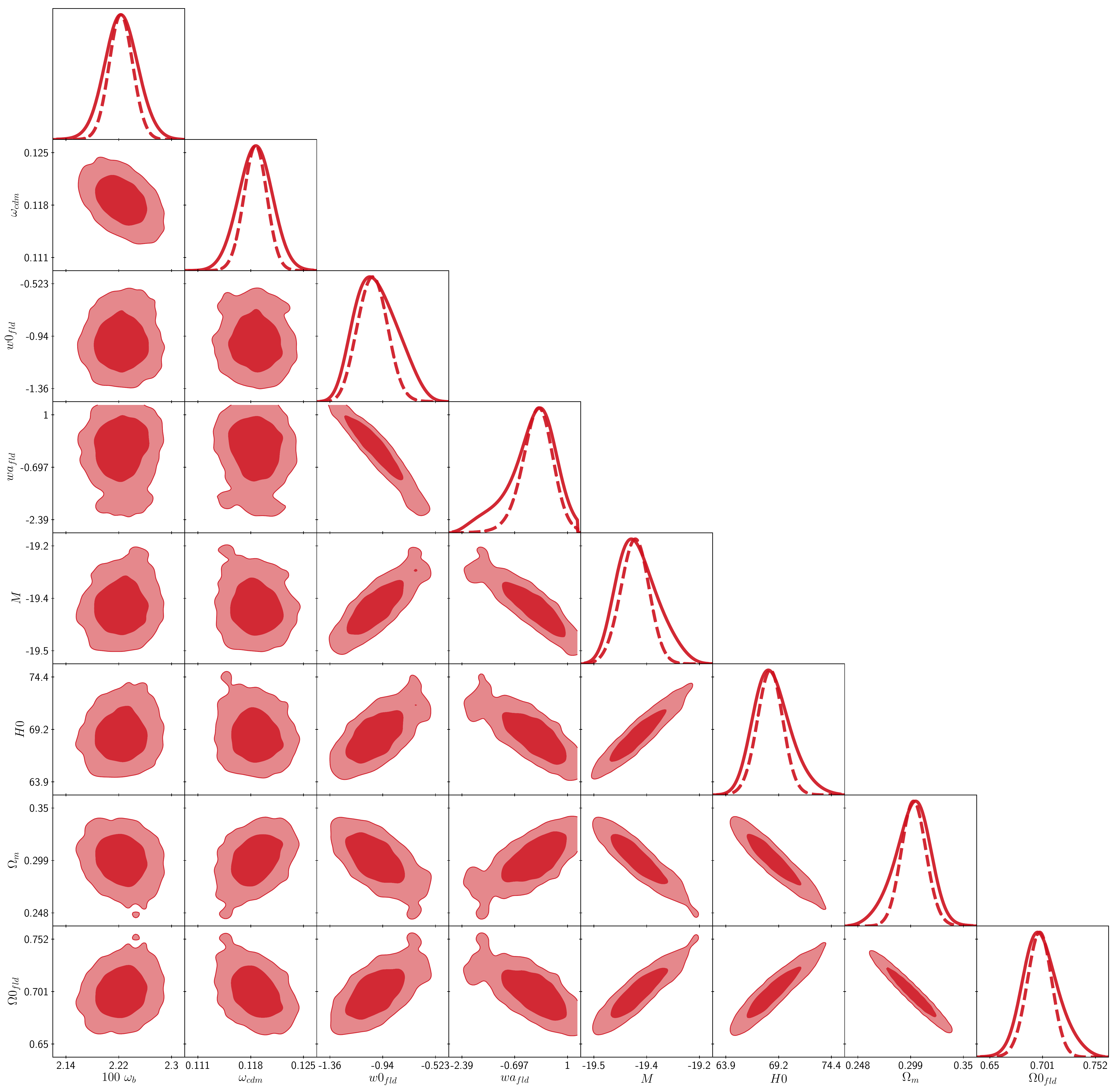}
               \includegraphics[width=0.42\textwidth,origin=c,angle=0]{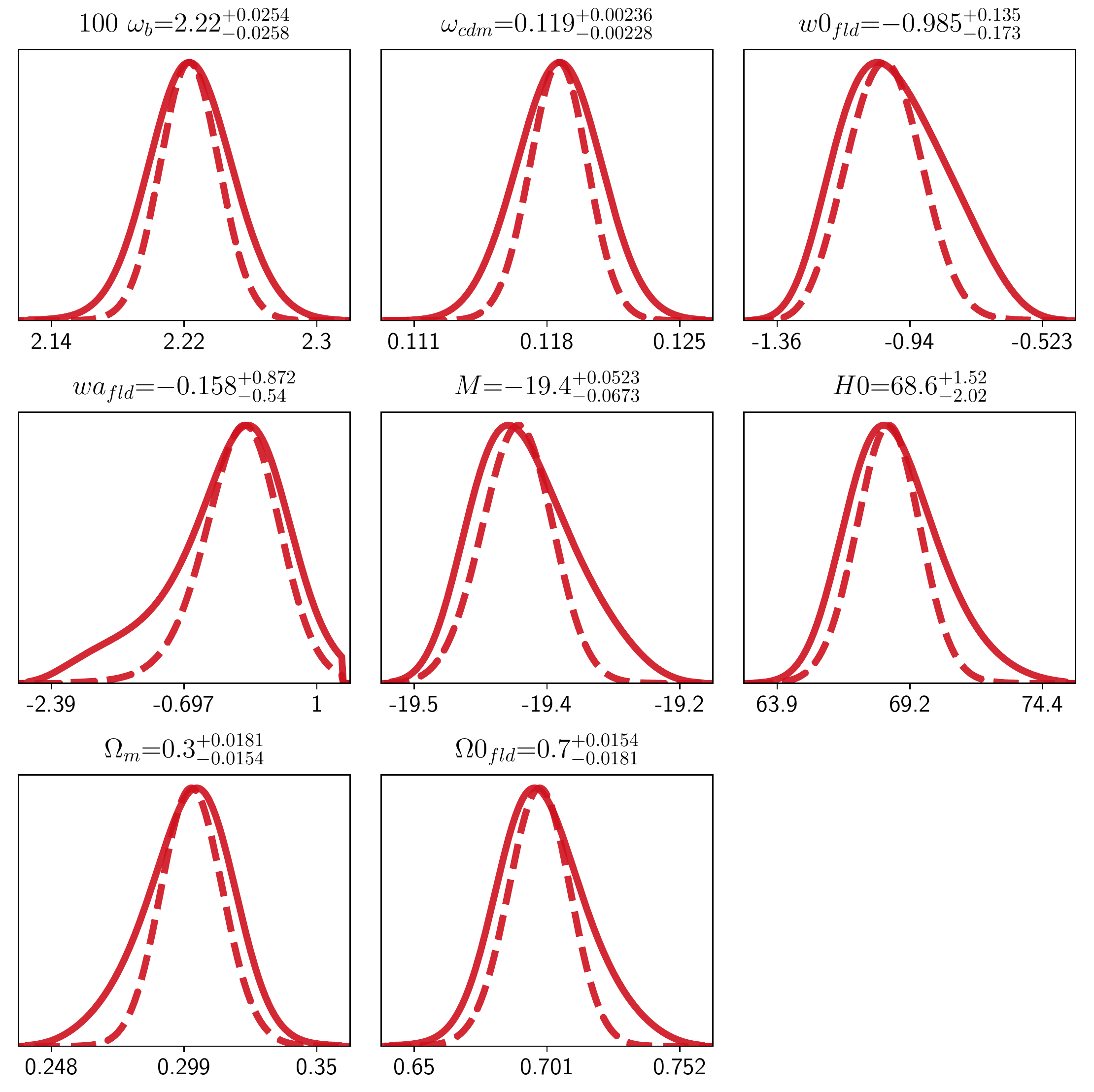}
    \caption{\textit{Left:} Confidence contours, \textit{Right:} Posterior, for CPL model using Deep Learning Pantheon sampler.}
    \label{fig:bayesian_pantheon_cpl}
\end{figure}

\begin{figure*}
    \centering
    \includegraphics[width=0.57\textwidth,origin=c,angle=0]{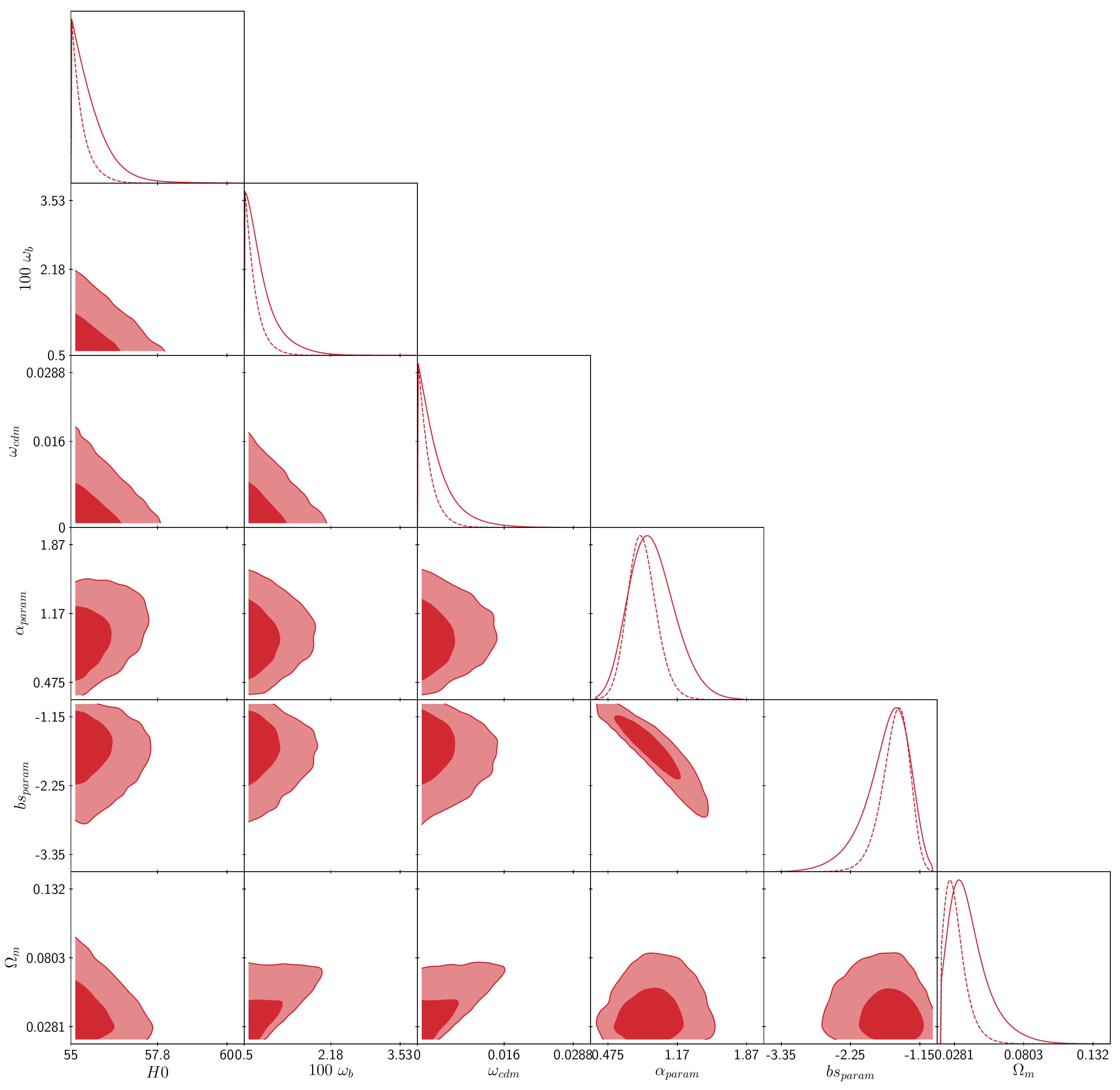}
       \includegraphics[width=0.37\textwidth,origin=c,angle=0]{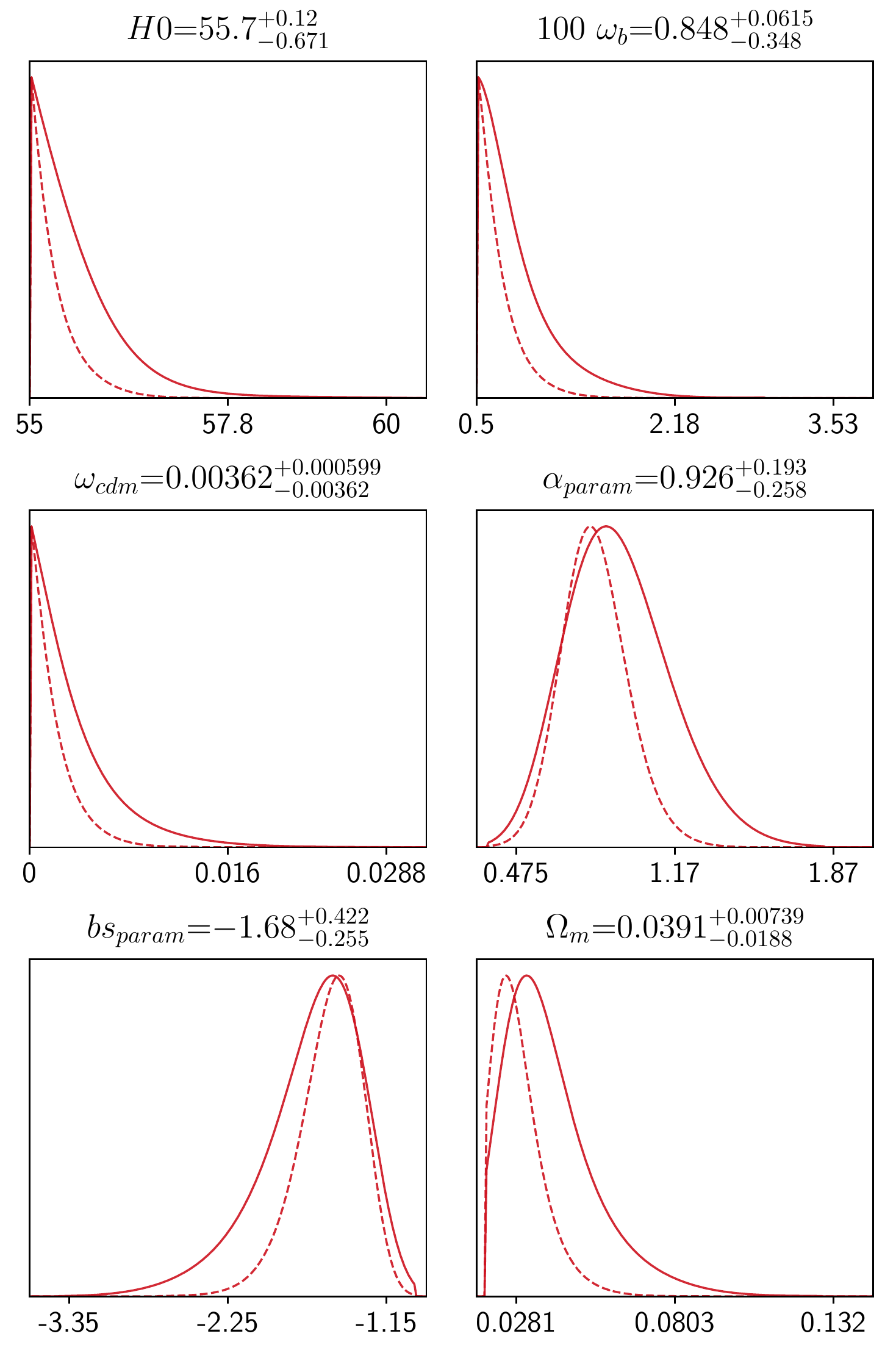}
    \caption{\textit{Left:} Confidence contours, \textit{Right:} Posterior, for Generalised Chaplygin Gas (GCG) model using Deep Learning Pantheon sampler.}
    \label{fig:bayesian_pantheon_gcg}
\end{figure*}

\begin{figure*}
    \centering
    \includegraphics[width=0.57\textwidth,origin=c,angle=0]{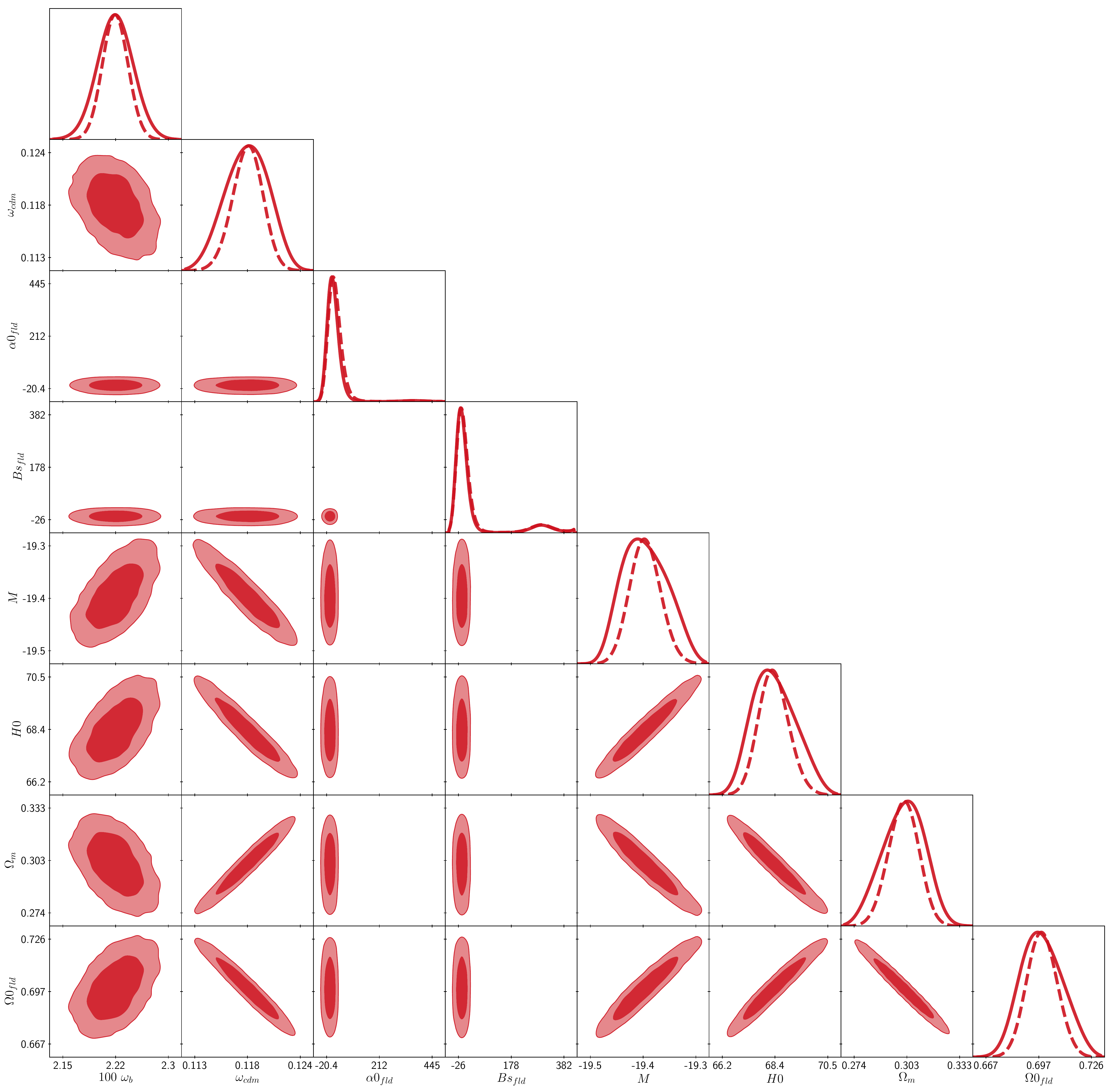}
                   \includegraphics[width=0.42\textwidth,origin=c,angle=0]{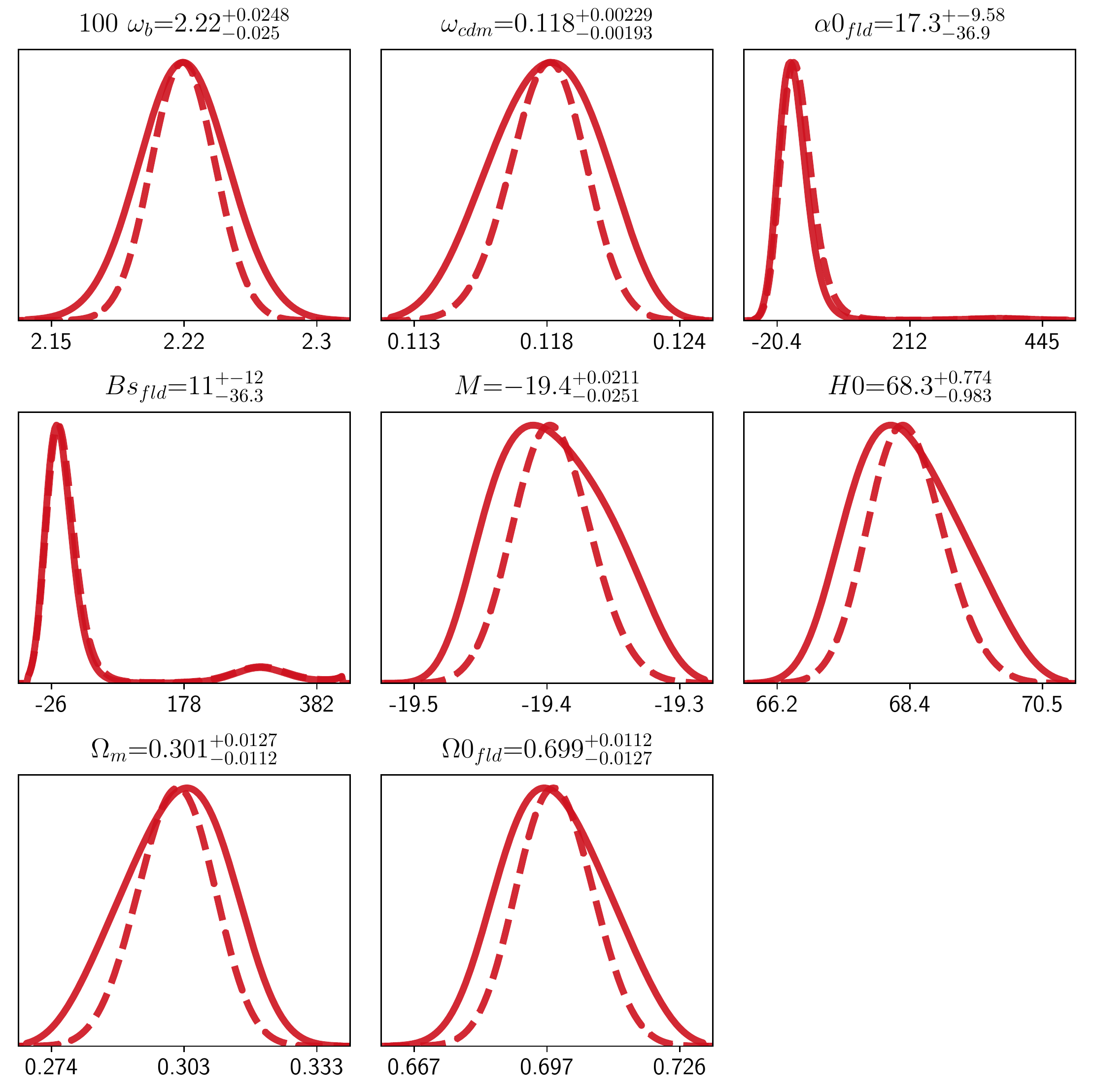}
    \caption{\textit{Left:} Confidence contours, \textit{Right:} Posterior, for Modified Chaplygin Gas (MCG) model using Deep Learning Pantheon sampler.}
    \label{fig:bayesian_pantheon_mcg}
\end{figure*}

In this section we present the parameter spaces using the Tanh activation function Eq.(\ref{eq:tanh}) according the results obtained in the latter sections. We proceed to compute the confidence regions for the models described in Sec.\ref{sec:background1} using our deep learning Pantheon sampler. In Figures \ref{fig:bayesian_pantheon_lcdm}-\ref{fig:bayesian_pantheon_cpl}-\ref{fig:bayesian_pantheon_gcg}-\ref{fig:bayesian_pantheon_mcg} are reported the confidence contours and posteriors for each cosmological model, respectively. Also we reported in Tables \ref{tab:lcdm}-\ref{tab:cpl}-\ref{tab:gcg}-\ref{tab:mcg} the best fit parameters within $95\%$ C.L. The results obtained from these analyses show a better $\chi^2$ using a deep learning trained sampler than about 1 unit in comparison to the pivot cosmological model. Furthermore, the tension at low redshifts seems to be reduced using this new sampler, alike the observational sampler, where a significant tension emerges at these $z$ scales.

On one other hand, model uncertainty is indispensable for the RNN. With model confidence at hand we can treat inputs and specific cases explicitly, e.g. in a scenario of classification a model can return a result with high uncertainty. In our RNN+BNN we used a specific Monte Carlo dropout which is supported by the Bayesian evidence. As an extension, several dropouts have been used in the literature \cite{DropoutRNN2} in order to obtain the predicted values for the network. Some of them have insensible values for the observational data, and the others have the same inconvenient but with the additional information that the models are uncertain about their predictions. In our case, with the Bayesian evidence developed, our selected dropout do not affect directly the form of the confidence contours in comparison to the activation function $A_{f}$. Each proposed $A_{f}$ can indeed change these confidence regions since they directly change their covariance matrix \cite{DropoutRNN}. Notice that our results using evidence are aligned with the reported in \cite{DropoutRNN2} with different dropouts. 

On the other hand, a strong case of support for our RNN+BNN network is the following: notice that our selected  $A_{f_{\text{Tanh}}}$ is bounded in comparison to the rest of activation functions. This is a hint to follow a physical interpretation over the trend of an observable: since we are using a sampler of SNeIa, we can expect that the observed differences in peak luminosities of SNeIa are very closely correlated with observed differences in the shapes of their light curves, i.e dimmer SNeIa decline more rapidly after maximum brightness, while brighter SNeIa decline more slowly. This behaviour is not given for ReLU and SELU, which makes them non-physical options of activation functions for our training according to the SNeIa trend. Moreover, Tanh and ELU can have a physical trend with the SNeIa, but ELU exhibit a higher error and weak evidence in comparison to Tanh. 

Notice that AIC is an estimate of a constant plus the relative distance between the unknown true likelihood function of the trained sampler and the fitted likelihood function of the cosmological model, so lower AIC means a model is considered to be closer to the truth. And the BIC is an estimate of a function of the posterior probability of a model being true, under a certain Bayesian setup, so that a lower BIC means that a model is considered to be more likely to be the true model. According to these ideas and our information criteria analyses, presented in Table \ref{table_bayesiancom}, SELU and ELU gives the lower AIC and BIC values in comparison to Tanh and ReLU. But, since SELU and ELU does not have a \text{physical} trend for our observable sample consider, as we explained above, Tanh is the only activation function with a better penalized-likelihood criteria. 

{\renewcommand{\tabcolsep}{6.mm}
{\renewcommand{\arraystretch}{0.5}
\begin{table}
\caption{Best fits values for $\Lambda$CDM model using Deep Learning Pantheon sampler.}
\begin{tabular}{|l|c|c|c|c|} 
 \hline 
Parameters & Best-fit & mean$\pm\sigma$ & 95\% lower & 95\% upper \\ \hline 
$M$ &$-19.43$ & $-19.43_{-0.011}^{+0.011}$ & $-19.45$ & $-19.4$ \\ 
$\Omega_{m }$ &$0.298$ & $0.299_{-0.023}^{+0.022}$ & $0.255$ & $0.344$ \\ 
$\Omega_{\Lambda }$ &$0.702$ & $0.701_{-0.022}^{+0.023}$ & $0.656$ & $0.745$ \\ 
\hline 
 \end{tabular} \label{tab:lcdm} \\ 
\end{table}}}

{\renewcommand{\tabcolsep}{6.mm}
{\renewcommand{\arraystretch}{0.5}
\begin{table}
\caption{Best fits values for CPL model using Deep Learning Pantheon sampler}
\begin{tabular}{|l|c|c|c|c|} 
 \hline 
Parameters & Best-fit & mean$\pm\sigma$ & 95\% lower & 95\% upper \\ \hline 
$100~\omega_{b }$ &$2.219$ & $2.223_{-0.026}^{+0.025}$ & $2.172$ & $2.275$ \\ 
$\omega_{cdm }$ &$0.119$ & $0.119_{-0.002}^{+0.002}$ & $0.114$ & $0.123$ \\ 
$w_0$ &$-1.024$ & $-0.985_{-0.17}^{+0.14}$ & $-1.284$ & $-0.641$ \\ 
$w_a$ &$0.064$ & $-0.158_{-0.54}^{+0.87}$ & $-0.003$ & $0.015$ \\ 
$M$ &$-19.4$ & $-19.39_{-0.067}^{+0.052}$ & $-19.52$ & $-19.26$ \\ 
$H_0$ &$68.41$ & $68.64_{-2}^{+1.5}$ & $65.03$ & $72.47$ \\ 
$\Omega_{m }$ &$0.301$ & $0.300_{-0.015}^{+0.018}$ & $0.265$ & $0.334$ \\ 
$\Omega_0$ &$0.699$ & $0.700_{-0.018}^{+0.015}$ & $0.667$ & $0.735$ \\ 
\hline 
 \end{tabular}  \label{tab:cpl}\\ 
\end{table}}}

{\renewcommand{\tabcolsep}{6.mm}
{\renewcommand{\arraystretch}{0.5}
\begin{table*}
\caption{Best fits values for GCG model using Deep Learning Pantheon sampler}
\begin{tabular}{|l|c|c|c|c|} 
 \hline 
Parameters & Best-fit & mean$\pm\sigma$ & 95\% lower & 95\% upper \\ \hline 
$H0$ &$55.07$ & $55.67_{-0.67}^{+0.12}$ & $55$ & $56.96$ \\ 
$100~\omega_{b }$ &$0.540$ & $0.848_{-0.35}^{+0.061}$ & $0.5$ & $1.548$ \\ 
$\omega_{cdm }$ &$0.001$ & $0.004_{-0.004}^{+0.001}$ & $1.499\times 10^{-7}$ & $0.011$ \\ 
$\alpha$ &$0.7628$ & $0.926_{-0.26}^{+0.19}$ & $0.4753$ & $1.396$ \\ 
$b_s$ &$-1.494$ & $-1.676_{-0.26}^{+0.42}$ & $-2.429$ & $-1.012$ \\ 
$\Omega_{m }$ &$0.018$ & $0.039_{-0.019}^{+0.007}$ & $0.016$ & $0.071$ \\ 
\hline 
 \end{tabular}  \label{tab:gcg} \\ 
\end{table*}}}

{\renewcommand{\tabcolsep}{6.mm}
{\renewcommand{\arraystretch}{0.5}
\begin{table*}
\caption{Best fits values for MCG model using Deep Learning Pantheon sampler}
\begin{tabular}{|l|c|c|c|c|} 
 \hline 
Parameters & Best-fit & mean$\pm\sigma$ & 95\% lower & 95\% upper \\ \hline 
$100~\omega_{b }$ &$2.224$ & $2.224_{-0.025}^{+0.025}$ & $2.175$ & $2.273$ \\ 
$\omega_{cdm }$ &$0.118$ & $0.118_{-0.002}^{+0.002}$ & $0.114$ & $0.122$ \\ 
$\alpha_0$ &$1.089$ & $17.35_{-37}^{+-9.6}$ & $0.002$ & $0.002$ \\ 
$B_s$ &$-1.901$ & $10.96_{-36}^{+-12}$ & $-0.013$ & $-0.013$ \\ 
$M$ &$-19.4$ & $-19.4_{-0.025}^{+0.021}$ & $-19.44$ & $-19.35$ \\ 
$H_0$ &$68.35$ & $68.34_{-0.98}^{+0.77}$ & $66.71$ & $70.14$ \\ 
$\Omega_{m }$ &$0.301$ & $0.301_{-0.011}^{+0.013}$ & $0.277$ & $0.323$ \\ 
$\Omega_0$ &$0.699$ & $0.699_{-0.013}^{+0.011}$ & $0.677$ & $0.722$ \\ 
\hline 
 \end{tabular}  \label{tab:mcg} \\ 
\end{table*}}}


\section{Conclusions} 
\label{sec:conclusions}

In this work we have trained a new neural network to parameterise efficiently dark energy EoS coming from different models: specifically we have trained  the standard $\Lambda$CDM, CPL  and
the unified dark fluids with Chaplygin gas characteristics. With these kind of parameterisations and their neural training, we reproduced the cosmic accelerated expansion currently observed. This novel technique
offers the combination of two architectures: (1) Recurrent Neural Networks (RNN) and (2) Bayesian Neural Networks (BNN), which we called BNN+RNN neural network. In comparison to other proposals reported in the DL literature \cite{DLmatrix}, with our proposed network we can train the observational data (which only consist on one input type $z$ and one output type $\mu(z)$), obtain trained data and compute their propagation error with a Bayesian architecture and the way we perform the network allows to the algorithm to be computational faster. Also, a direct physical trend related to the activation functions can be followed in the case of homogeneous sampler as SNeIa.
 In these kind of processes, the quality in the density of data points is important. 
In such case, our choice of $A_f$ can be viewed as a \textit{choice of a prior} since it is relevant in regions where the training data is sparse (or even non-existent) and the resulting trained DL sampler explicitly depends of this choice.
Therefore,
the described scheme is tested with SNeIa data available today. Finally, 
our new method has a wide range of applicability to many problems in cosmology, 
for example to explore new constrains on cosmographic parameters. 
This latter is a working project currently in progress.


\acknowledgments

CE-R acknowledges the \textit{Royal Astronomical Society} as FRAS 10147. CE-R and MACQ are supported by \textit{PAPIIT} Project IA100220 and ICN-UNAM projects. 
SC is supported by \textit{Istituto Nazionale di Fisica Nucleare} (INFN), \textit{Iniziative Specifiche} QGSKY and MOONLIGHT2. This article is also based upon work from COST action CA18108, supported by COST (European Cooperation in Science and Technology). The authors thank the anonymous reviewers whose comments have improved this manuscript.


\end{document}